\documentstyle[aps]{revtex}
\begin{document}
\draft
\title{Low temperature thermal conductivity of Zn-doped YBCO: evidence for
impurity-induced electronic bound states}
\author{K. Behnia, H. Aubin}
\address{Laboratoire de Physique des Solides (associ\'e au CNRS),\\
Universit\'e Paris-Sud, 91405 Orsay, France }
\author{L. Taillefer, R. Gagnon}
\address{Physics Department, McGill University, Montr\'eal, Qu\'ebec,Canada H3A 2T8}
\date{\today }
\maketitle

\begin{abstract}
The thermal conductivity of Zn-doped YBCO crystals was studied at low
temperature (0.15 K%
\mbox{$<$}
T%
\mbox{$<$}
0.8 K) for different concentrations of Zn impurities. A small amount of Zn
induces a dramatic decrease in the non-linear component of the
low-temperature thermal conductivity. Moreover, the magnitude of the linear
component (obtained by extrapolating the data to T=0) is found to depend on
Zn concentration. After an initial decrease, this linear term, associated
with the electronic contribution to the conductivity, increases with
increasing Zn dopage. Such an increase is consistent with the introduction
of low-energy excitations by Zn impurities as expected for a $d_{x^2-y^2}$
superconducting state in contrast to an anisotropic s-wave gap. The results
are compared to quantitative predictions of available theoretical models.
\end{abstract}



The identification of the symmetry of the superconducting order parameter in
the high-T$_c$ cuprates is a main task of experimental and theoretical
research in the field. Evidence for nodes in the superconducting gap of YBCO
has been accumulated during recent years. Among others, one can mention
experimental data resulting from magnetic penetration depth\cite{zhang},
angular resolved-photoemission\cite{shen}, nuclear spin relaxation rate \cite
{ishida}, low-temperature thermal conductivity\cite{bredl,behn} measurements
and most recently the unusual dependence of thermal conductivity on the
orientation of a magnetic field rotating in the basal plane\cite{yu,salamon}
. An order parameter with a $d_{x^2-y^2}$ symmetry \cite{scal} is the most
promising candidate to account for the anisotropic gap implied by these
experiments. But, as was pointed out recently \cite{chakra}, an
anisotropic s-wave gap, with deep minima at the same angular positions as
the nodes of the $d_{x^2-y^2}$ state, would be compatible with this set of
experiments.

One way to distinguish between these two possible states is to exploit the
sharp contrast in their response to non-magnetic scattering. Contrary to an
anisotropic s-wave state, where the introduction of non-magnetic scatterers
tends to eliminate node-like deep angular minima in the gap, the
single-particle density of states in a d-wave state increases with
increasing concentration of such impurities\cite{borkow,fehren}. In this
paper, we present the results of a systematic study of the low-temperature
thermal conductivity of $YBa_2(Cu_{1-x}Zn_x)_3O_{6.95}$ single crystals
which shows that for 0.006%
\mbox{$<$}
x%
\mbox{$<$}
0.022, the quasi-particle thermal conductivity increases with increasing x ,
providing new support in favor of d-wave symmetry.

YBCO single-crystals were prepared by a self-flux method as described
elsewhere\cite{gag}. Zn content was determined using the known dependence of
T$_c$. Measurements were performed in a dilution refrigerator using a
one-heater-two thermometer steady-state method. One difficulty in obtaining
the absolute value of the thermal conductivity and in comparing different
samples, is the precise determination of the geometric factor. Sample
resistance at room temperature was used to check the value obtained through
the direct measurement of the dimensions. The uncertainty on the magnitude
of the geometric factor thus obtained is estimated to be about 20\%.

Fig. 1 shows the results for three samples with different concentrations of
Zn impurities. To simplify the separation of electronic and lattice
components of the thermal conductivity, $\kappa $ , we plot $\kappa $/T against T%
$^2$. The first striking feature in this figure is the sudden change of
behavior of $\kappa $/T with introduction of a very small amount of Zn. The
extrapolation of data to zero temperature yields a finite linear term for
all the samples. Below 0.35 K, a cubic term can be extracted using a $\kappa
=aT+bT^3$ fit. Fig. 2 presents the variation of the linear and cubic terms
with the concentration in Zn. The magnitude of both the linear and cubic
terms in the pure crystal is lower than what we had previously obtained for
untwinned (and undoped) crystals\cite{behn}, suggesting that twin boundaries
limit the maximum mean-free-path of heat carriers.

The first question to address before using thermal conductivity as a probe
of electronic excitations concerns the extraction of the lattice
contribution. The phonon thermal conductivity is expected to show a T$^3$
behavior at low temperature, if the mean-free-path of the phonons is limited
by the finite size of the crystal. Using the low-temperature specific-heat
data \cite{collo}, and the sound velocity\cite{zou}, one can estimate the
magnitude of the lattice cubic term imposed by the smallest dimension of the
crystal. Taking this to be between 50 and 100$\mu m$, i.e. the typical
thickness of our crystals, one obtains a T$^3$ term of about 2 to 4 mW/ K$^4$
cm. This is close to the magnitude of the cubic term obtained for x 
\mbox{$>$}
0.002 but about half of the corresponding value for the undoped crystal and
only a third of the value for the untwinned crystals. Thus, as we discuss
below, one cannot exclude an 
electronic component to the non-linear thermal
conductivity.

One complication in the interpretation of our data is the possible increase
in the concentration of twin domains with Zn-doping. Given the effect of
twin boundaries on the mean-free-path of heat carriers, both phonon and
electronic contributions could be affected. It is not excluded that the
initial decrease in the magnitude of both the linear and cubic terms for 
0
\mbox{$<$}
x
\mbox{$<$}
0.002 is a result of a sudden change in the density of twin boundaries.

The linear term, beyond the initial decrease, increases monotonically for x 
\mbox{$>$}
0.006. Such an increase suggests an enhancement in available electronic
excitations which would exceed the reduction of the quasi-particle
mean-free-path due to impurities. The increase in the very-low temperature
electronic density of states with Zn doping is the main qualitative outcome
of our experiment. Recently, theoretical calculations showed that the
response to doping by non-magnetic impurities in a $d_{x^2-y^2}$
superconducting state is qualitatively different from that of an anisotropic
s-wave superconductor\cite{borkow,fehren}. The latter was suggested as an
alternative to the d-wave candidate in order to account for experimental
results indicating a finite density of states at zero temperature. While in
a d-wave superconductor, the pair-breaking effects of non-magnetic
impurities would lead to an increasing zero-energy density of states, in the
s-wave counterpart, quite to the contrary, it would reduce the gapless
behavior. Therefore our results argue strongly in favor of the d-wave case.
One complication, however, arises from the magnetic moment induced by Zn
impurities in the $CuO_2$ plane as detected by nuclear magnetic resonance
measurements above T$_c$\cite{alloul}. A gapless behavior due to the
appearance of spin scattering by the induced local moments would thus be
expected even in the case of s-wave superconductivity. A crude estimate of
spin-flip scattering in Zn- doped YBCO, however, suggests that the
non-magnetic scattering dominates the total elastic rate \cite{borkow} and
the global response of a hypothetical anisotropic s-wave gap to Zn doping
would continue to be a decrease in density of low-energy excitations in
contrast to our experimental data.

Next, we turn to compare our data with the results of several recent
theoretical studies\cite{lee,hirsch,sun,graf} on the quasi-particle
transport in a d-wave superconductor in presence of resonant scattering by
non-magnetic impurities. The picture developed in these works leads to a
number of predictions on the low-temperature behavior of the electronic thermal
conductivity, $\kappa _e$ and the 
low-frequency microwave conductivity, $\sigma $%
: i) Both $\sigma $ and $\kappa _e$/T should tend to universal values at
zero-temperature which are independent of the relaxation rate, their ratio
being equal to the Sommerfeld value L$_0$; ii) At finite temperatures above
the universal regime but low enough to neglect the inelastic scattering, $%
\sigma $ should follow a quadratic and $\kappa _e$ a cubic temperature
dependence.

There remains a discrepancy
between this model and the available experimental data.
Surface resistance measurements\cite{zhang}, yielding the temperature
dependence of microwave conductivity, did not detect any quadratic behavior
down to 2K. On the other hand, a $linear$ extrapolation of data to T=0 would
reproduce a value close to the expected theoretical magnitude for $\sigma _0$%
\cite{zhang}. In the case of thermal conductivity, the universal value for
the residual $\kappa /T$ is $\frac{\omega _p^2k_b^2h}{4e^2\Delta _0}$ which,
taking reasonable values for the 
plasma frequency and for the gap ( $\omega _p=1.4eV
$  and $\Delta _0=2.14k_bT_c),$ would yield 0.11- 0.14 mW/Kcm\cite{graf}.
The experimental value obtained for detwinned undoped crystals by
extrapolating to T=0 the data obtained down to 0.15 K is three times larger
(0.4mW/Kcm\cite{behn}). However, the energy scale characteristic of the
universal linear regime for the electronic thermal conductivity depends on
the scattering rate (i.e. impurity concentration) and on the scattering
phase shift\cite{graf}. Therefore, the universal regime, pushed down to very
low temperatures in the pure case, may have not been attained in our
experiments on the undoped crystals. Such an hypothesis can provide a
possible explanation for the sudden decrease in the magnitude of the cubic
term between x=0.001 and x=0.002 in our results (Fig. 2b). Indeed, this can
be attributed to a crossover between two different regimes. In such a
scenario, for x 
\mbox{$<$}
0.002, the low-temperature regime is not reached in the temperature range
investigated (T/T$_c$ 
\mbox{$>$}
2 10$^{\text{-3}}$) so that the electronic contribution is essentially
cubic. This would explain the large value of the cubic term for the purest
samples which, as discussed above, given the estimates of
phonon-mean-free-path, is perhaps not of exclusively lattice origin. For x%
\mbox{$>$}
0.002 on the other hand,  the cubic term would entirely be due to the
phonons and the electronic contribution would become linear. As mentioned
above, there is also the more prosaic interpretation associated with
a possible change in the density of twin boundaries to explain the sudden
decrease in the cubic term.

According to the theoretical model discussed above, in the unitary limit,
the residual $\kappa _e$/T should increase monotonically with impurity
concentration\cite{sun}. The experimental data (Fig. 2 b) does not fit to
this scheme. At low concentrations, we detect a decrease in the linear term,
slightly larger than the experimental uncertainty and then, at higher
concentrations, a considerable increase, which exceeds what is expected by
ref. 18. Once again the scattering of quasi-particles by twin boundaries and
the change in the density of the latter with Zn concentration complicates a
direct comparison of data with theory.

In conclusion, our results provide evidence for introduction of low-energy
excitations by Zn impurities in YBCO, as expected for a d-wave
superconductor. However, to test the quantitative predictions of the current
theoretical model on the variation of the residual quasi-particle thermal
conductivity with impurity concentration, measurements on detwinned samples,
currently under way, are necessary.

We acknowledge useful discussions with K. Maki, M.-T. Beal-Monod, P.
Hirschfeld, M. J. Graf and J.A. Sauls.

\begin{figure}
\caption{Temperature dependence of thermal conductivity,
divided by temperature, for various concentrations of Zn.}
\label{fig. 1}
\end{figure}

\begin{figure}
\caption{Dependence of the linear (a) and cubic (b) terms of low-temperature 
thermal conductivity. a and b are extracted from the best fit to 
$\kappa =aT+bT^3$ for T %
 0.35 K.}
\label{fig. 2}
\end{figure}

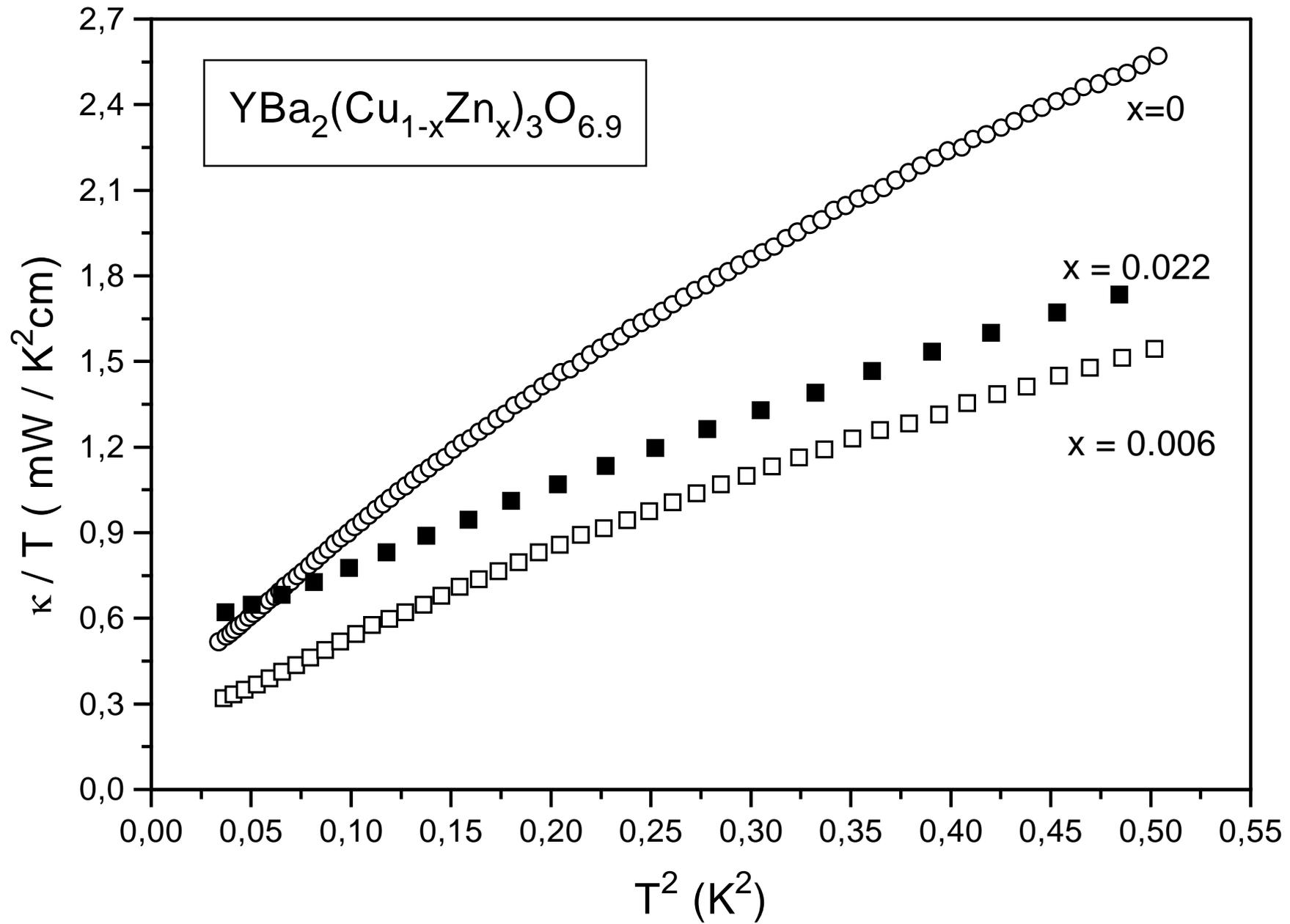
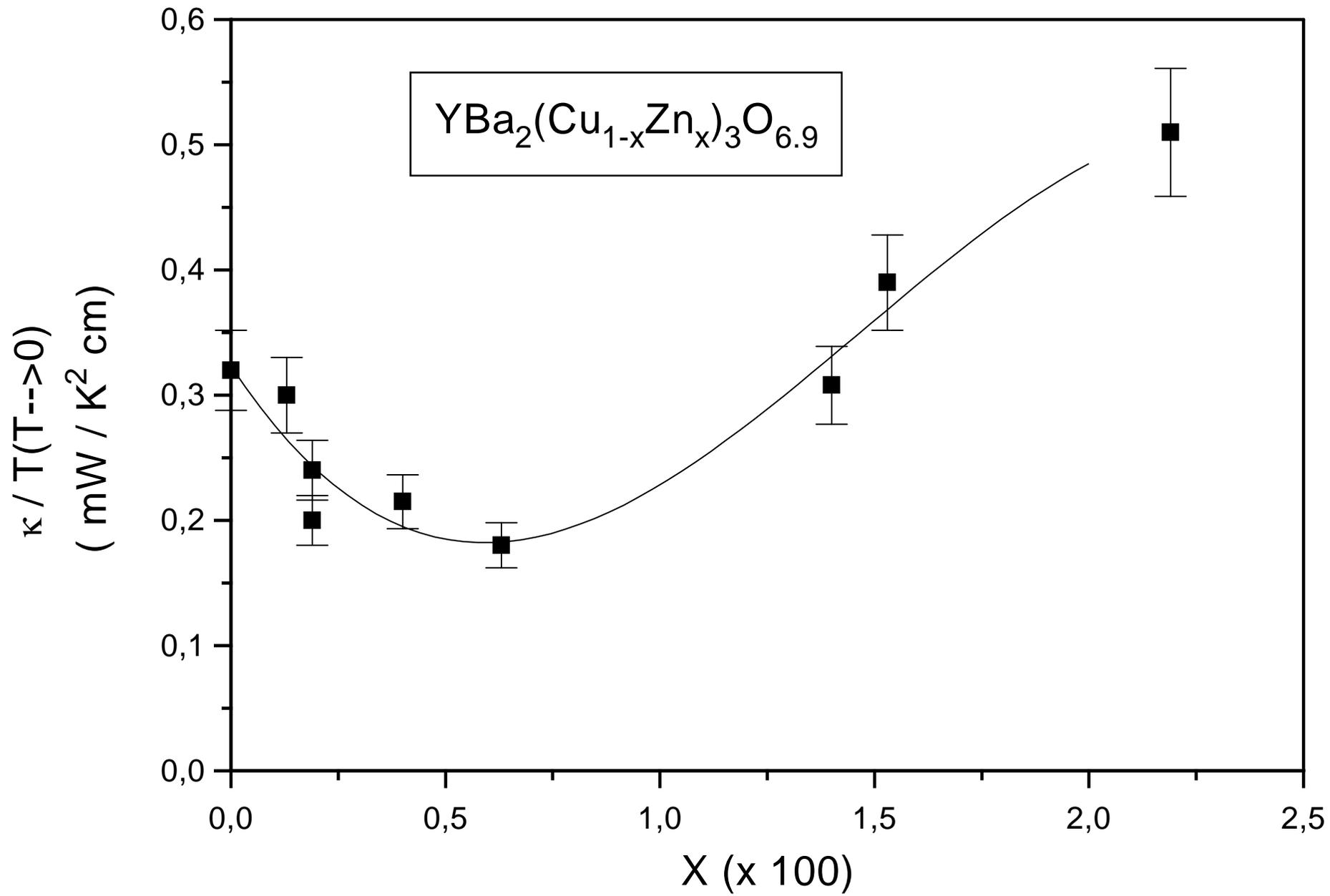
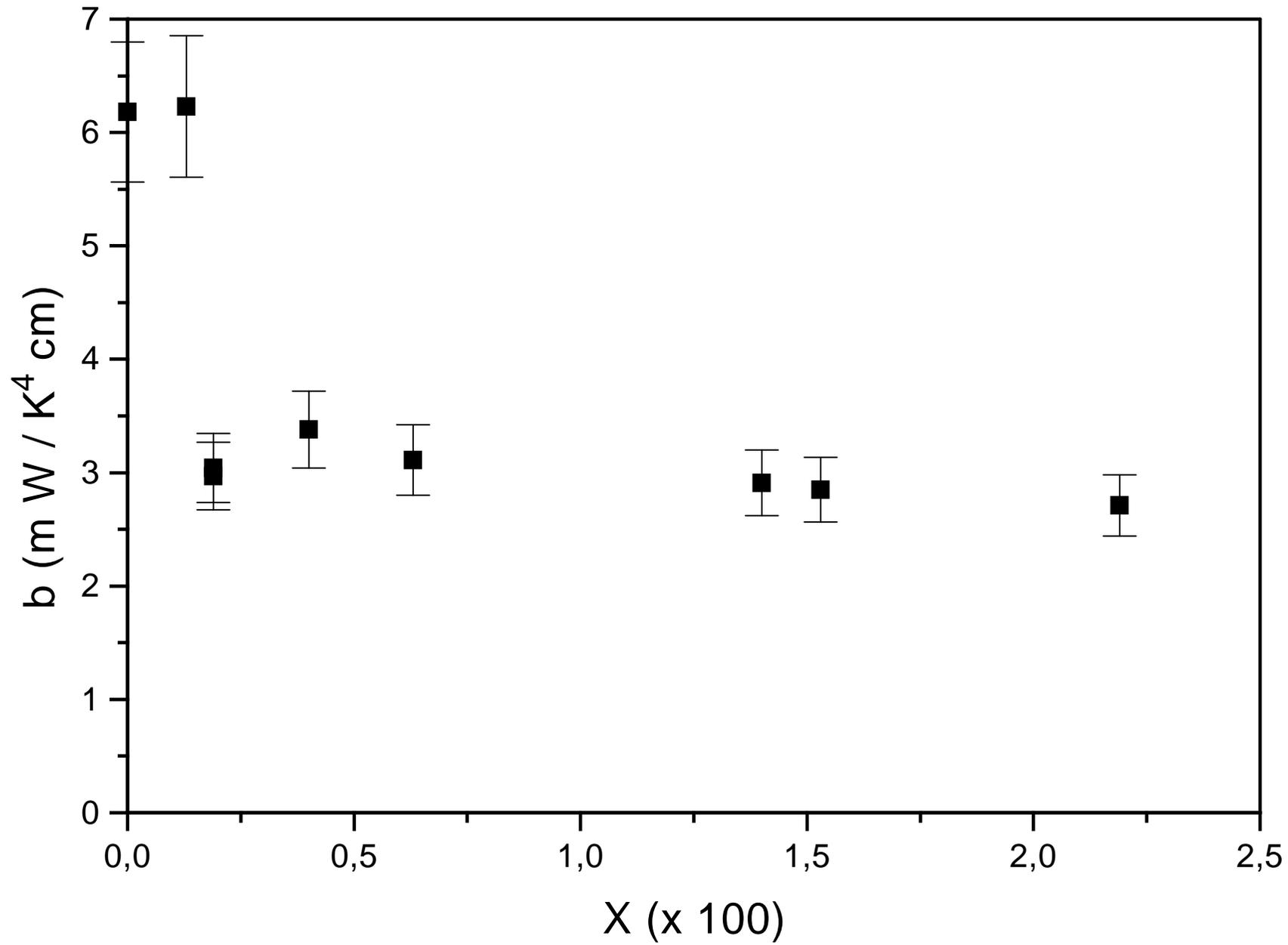
\end{document}